\documentclass[12pt,preprint]{aastex}

\slugcomment{accepted by $ApJ$ $Letters$ on Feb. 1, 2007}

\shorttitle{Statistics of Highly-Irradiated Exoplanets}
\shortauthors{Hubbard et al.}

\begin{document}

\title{A Mass Function Constraint on Extrasolar Giant Planet Evaporation Rates}

\author{W. B. Hubbard and M. F. Hattori}
\affil{Lunar and Planetary Laboratory, The University of Arizona,
    Tucson, AZ 85721-0092}
\email{hubbard@lpl.arizona.edu, makih@lpl.arizona.edu}

\and

\author{A. Burrows and I. Hubeny}
\affil{Department of Astronomy, The University of Arizona,
    Tucson, AZ 85721-0065}
\email{burrows@as.arizona.edu, hubeny@as.arizona.edu}

\begin{abstract}
The observed mass function for all known extrasolar
giant planets (EGPs) varies approximately
as $M^{-1}$ for mass $M$ between $\sim0.2$ Jupiter masses ($M_J$)
and $\sim5$ $M_J$.  In order to study evaporation effects for
highly-irradiated EGPs in this mass range, we have constructed
an observational mass function for a subset of EGPs in the same
mass range but with orbital radii $<0.07$ AU.  Surprisingly, the
mass function for such highly-irradiated EGPs agrees quantitatively with
the $M^{-1}$ law, implying that the mass function for EGPs is
preserved despite migration to small orbital radii. Unless there
is a remarkable compensation of mass-dependent orbital migration
for mass-dependent evaporation, this result places a constraint
on orbital migration models and rules out the most extreme mass
loss rates in the literature. A theory that predicts more moderate
mass loss gives a mass function that is closer to observed
statistics but still disagrees for $M < 1$ $M_J$.
\end{abstract}


\section{Introduction}

The rate of mass loss during evolution of highly-irradiated EGPs
has important implications for our understanding of the
origin of these unanticipated objects.  The principal purpose of this {\it Letter}
is to show that enough such objects have now been detected to permit
construction of a mass function for highly-irradiated EGPs, in the mass range where
strong evaporation effects had been predicted \citep{bar04}. The derived
mass function shows no evidence for mass loss; its resemblance to the
mass function for all observed EGPs may instead be a constraint on migration
theories.

Our analysis \citep{hub06} is based on a single dimensionless parameter $\epsilon$
which represents the efficiency of the mass loss process:
\begin{equation}
\epsilon = \Phi E_{B,mod}/S_* ,
\end{equation}
where $\Phi$ is the flux of escaping molecules (number per unit area per unit time),
$E_{B,mod}$ is the gravitational binding energy of a molecule to the planet, as
modified by the tidal potential, and $S_*$ is the bolometric stellar flux (energy
per unit area per unit time) at the planet's orbital distance, for a main-sequence star
of solar mass and age.  \citet{hub06} calculate
the EGP mass-loss rate as a function
of time by coupling this parameter to
evolutionary models for EGPs orbiting solar-mass stars at orbital radii in the range
$0.023$ AU to $0.057$ AU.  The calculations of \citet{bar04} can be reproduced by
setting $\epsilon = 10^{-4}$.  A model scaled to the calculations of
\citet{wat81}, as well as to the lower limit considered by \citet{bar06},
corresponds to $\epsilon = 10^{-6}$.  In this paper, we use the
suite of models of \citet{hub06} to investigate the cumulative effect of mass
loss on an initial mass function for highly-irradiated EGPs.

\section{Initial Mass Function}

We assume an initial mass function $f(M,0)$ defined as follows.  The
database of EGPs known in 2005 \citep{mar05} has a mass function (the mass $M$
is multiplied in most cases by the unknown sine of the orbit inclination $i$)
corresponding to
\begin{equation}
f(M,0) \equiv dN/dM \propto M^{-1}.
\end{equation}
Our independent fit to a slightly different database, the table of $M \sin i$
and orbital semimajor axis $a$ for reported EGPs as given by \citet{sch06} in
mid-2006, gives a similar result,
$dN/dM \propto M^{-1.19}$.  As discussed by \citet{bur01} and \citet{mar05}, when one corrects
the underlying
function $dN/dM$ for random orbit inclinations, the index is essentially unchanged,
and for an index of $-1$ it is exactly unchanged.

In this paper, we define the initial mass function (IMF) for highly-irradiated EGPs
somewhat differently than the IMF for stars, the latter referring to the stellar birth
function. Since the majority ($\sim75$\%) of detected EGPs have orbital radii $> 0.07$
AU, they cannot have suffered significant mass loss from atmospheric escape over
their lifetime.  We refer to the latter class as ``field'' EGPs.  Our hypothesis is that
highly-irradiated EGPs are not formed {\it in situ}, but are remnants of field
EGPs that have migrated inward to small orbital radii during the $\sim10^6$ to $10^7$
years that the star's initial planet-forming nebula persists.
Let the IMF denote the initial mass function for highly-irradiated
EGPs at age $t\sim10^6$ to $10^7$ yr in a mass range
$\sim0.2$ $M_J < M <\sim5$ $M_J$, when these EGPs start their atmospheric erosion.
This IMF may well differ from the observed mass function
for field EGPs, since the
migration mechanism could have a mass
dependence.  One prediction of the IMF for highly-irradiated EGPs
\citep{del05} suggests a depletion of planets with $M>4$ $M_J$ but that
lower-mass EGPs migrate inward readily (see also \citet{tri98} and \citet{tri02}).
In this paper we make the provisional assumption that the IMF has the same index
as the mass function for field EGPs.

Figure 1 shows our assumed IMF $f(M,0)$ (heavy solid line),
arbitrarily normalized to unity at $M=1$ $M_J$.  We assume that EGPs
born at $a\sim$ several to several $\times 10$ AU are deposited at smaller orbital radii
$a\sim$ few$\times 10^{-2}$ AU within the
first few$\times 10^6$ years of the parent star's lifetime. 
To map the IMF onto a time-dependent ensemble of eroding EGPs, we fix the distance
from the star at one of the four standard distances studied in \citet{hub06}, ranging
from 0.023 to 0.057 AU.  We then randomly choose an exoplanet of
initial mass $M_0$ from the IMF and allow it to lose mass as a function of time $t$
using either the \citet{wat81} or the \citet{bar04} prescription.

It is necessary to impose a low-mass cutoff to the IMF.  As shown by
\citet{hub06}, hydrogen-rich EGPs with masses $\le0.2$ $M_J$ and initial entropies corresponding
to isolated EGPs at $\sim10^6$ years of age, have such large radii that their
atmospheres are tidally unbound (independent of the mass-loss rate), and so we
restrict our analysis to $M_0 \ge 0.2$ $M_J$.

\section{Mass Loss Models}

We have synthesized time-dependent mass functions $f(M,t)$
for EGPs at the four orbital radii investigated by \citet{hub06}.  
Figure 1 shows resulting mass functions
at $t=5$ Gyr after simultaneous mass loss and evolution of the EGP.
In all cases, the mass functions are normalized to unity at $M = 1$ $M_J$.  For
fixed total initial mass of the ensemble, all subsequent mass
functions would plot below the IMF due to mass loss, with the
strongest deviations at the lowest mass.  However, we renormalize
the theoretical curves to 1 $M_J$ for comparison
with the observed mass function data points.

\subsection{Lammer Model}

We denote by ``Lammer Model'' the evaporation theory of \citet{lam03}
as incorporated in the predictions of \citet{bar04}.
We do not separately simulate the
implications of more moderate mass-loss rates investigated in more
recent publications \citep{bar06,ali06} because these are bounded by
the \citet{bar04} rates and the Watson model.
In agreement with the findings of \citet{hub06}, 5-Gyr mass
functions corresponding to the Lammer model
show the strongest effect of mass loss.  At the smallest
orbital radius investigated, $a=0.023$ AU, the predicted 5-Gyr mass
function (solid curve) has a positive slope for the entire
mass interval plotted in Fig. 1, while the IMF has a negative slope.
This behavior occurs because mass loss biased to the lowest-mass
EGPs rapidly depletes their larger initial numbers.  Indeed,
the mass function plotted is almost entirely populated by remnants of
EGPs with initial masses $M_0 > 3-4$ $M_J$
(see Fig. 7 of \citet{hub06}).  For the largest orbital radii
investigated, $a=0.057$ AU, after 5 Gyr
there {\it is} a peak in the mass distribution at $M \sim 2.3$ $M_J$.

\subsection{Watson Model}

In contrast, the Watson model predicts essentially no deviation of
the mass function from the IMF
for $M > 1$ $M_J$.  However, as we see from the curves in
Fig. 1, for $M < 1$ $M_J$ the Watson model predicts significant
deviations from the IMF, with a maximum at about one Saturn mass
for $a=0.057$ AU and at even higher masses for smaller orbital radii.

\section{Observed Mass Function}

The database from which we construct the mass function for highly-irradiated EGPs
comprises $\sim40$ objects (Table 1).  These objects were selected from the list
of all reported EGPs \citep{sch06} according to the criteria $0.2$ $M_J \le M \sin i \le 5$ $M_J$ and
$a \le 0.07$ AU.  We selected a central mass bin of width $0.3$ $M_J$, according to
$0.9$ $M_J \le M \sin i < 1.2$ $M_J$, containing four objects (Table 1), with $dN/dM$ to be
normalized to unity at this bin.  Bins were chosen to be of equal width to the central bin, except
for the smallest-mass bin $0.2$ $M_J \le M \sin i < 0.3$ $M_J$, which is of width $0.1$ $M_J$,
because of the cutoff at $0.2$ $M_J$, and for the largest-mass bins,  $1.5$ $M_J \le M \sin i < 3$ $M_J$
and  $3.0$ $M_J \le M \sin i < 4.5$ $M_J$, which are five times wider because of the small number
of high-mass objects in the sample.

The resulting mass function is plotted in Fig. 2, with error bars determined by Poisson statistics.
Also shown in Fig. 2 is the average orbital radius (in AU) of objects in the bin (number to the
left of the error bar), and the average age (in Gyr) of the host stars in the bin (number to the
right of the error bar).

It is remarkable that the mass function for these highly-irradiated objects is in agreement with
the field IMF (see \citet{mar00} for a discussion of selection bias).

\section{Mass Function With $\sin i$ Factor}

For comparison of the mass-loss models with observed statistics, we require a transformation of the
theoretical mass function to a function of the new independent variable $M \sin i$.
Let $f(M,t)$ be the mass function of an eroded ensemble of EGPs after time $t$.  We write
\begin{equation}
f(M,t)=f(q/\sin i,t),
\end{equation}
where the new independent variable $q=M \sin i$.  Now holding the observed quantity $q$
fixed, we average the ensemble of eroded EGPs over all
solid angles of presentation of their
orbits to the observer $d\Omega$, according to
\begin{equation}
g(q,t) = \int {{d\Omega} \over {4\pi}} f(q/\sin i,t) = \int_0^1 d\mu f(q/\sqrt{1-\mu^2},t),
\end{equation}
where $d\mu=\sin i di$.  As would be expected, averaging over $\sin i$ moves the maxima of the
mass functions to slightly smaller masses.  The resulting functions $g(q,t=5$ Gyr$)$ for the Watson
model (curves to the left of $M \sin i = 1$ $M_J$) and the Lammer model (curves to the right of $M \sin i = 1$ $M_J$)
are plotted in Fig. 2.

\section{Conclusions}

Our results strongly suggest that the
mechanism for transporting EGPs to small orbital radii does not
significantly change their mass function in the mass range considered here,
and that unless a remarkable compensation of evaporation for migration has
occurred, we have no evidence for evaporation effects.  It is difficult
to envision such a compensation.  For example, to fit
the observed mass function using the Lammer evaporation model, we would need
to start with an IMF with an index $\sim -3.6$, but the fit would be highly
sensitive to the ages of the objects and would rapidly change with time.
Instead, a parsimonious interpretation of the data would be that
mass loss does not play a dominant role in even the hottest EGPs
with $M > 0.2$ $M_J$.
If we take the observed numbers of EGPs in the two innermost
mass bins to be statistically valid, even the more moderate
Watson model still disagrees
with the data, but certainly less decisively than does the Lammer model.
However, our arguments seem to point in the direction of requiring the mass-loss
parameter $\epsilon$ to be even smaller than $10^{-6}$. All
published theories give values higher than this number; for example \citet{tia05}
and \citet{yel06} have $10^{-6} < \epsilon < 10^{-4}$.

We must mention some caveats.  
Comparison of theory with the observational database is complicated by the fact that
a majority of the observed objects in the bins
$0.6$ $M_J \le M \sin i \le 1.2$ $M_J$ are
transiting EGPs (indicated by an asterisk in Table 1), and thus the masses
in these bins are true masses without the $\sin i$ ambiguity. A more consistent
mass function can be derived once we have enough transiting EGPs to construct a
statistically significant mass function
for them alone.  Also, we have not included ``Neptunes'' in
our analysis, although some highly-irradiated ``Neptunes'' do exist
\citep{but04,mca04,san04,lov06}.  Our assumed sharp mass cutoff at $0.2$ $M_J$ is actually
an approximation to a somewhat fuzzy boundary whose details depend greatly
on the exact model for tidal effects and on the initial entropy of
such low mass hydrogen-rich objects.  Our previous analysis \citep{hub06}
confirms that such ``Neptunes'' cannot be Jupiter-like EGPs (i.e., predominantly
hydrogen), but must have large fractions of elements with $Z>2$.
Whether nature could ever form hydrogen-rich
low-mass (Neptune) EGPs is highly debatable.  We eagerly await observations
of transits by ``Neptunes''. 

As more transiting EGPs are detected (and these
will be predominantly highly-irradiated ones), a more definitive 
statistical test of orbital-migration and mass-loss theories will be possible.

\acknowledgments

This study was supported in part by NASA Grant NAG5-13775 (PGG),
NASA grant NNG04GL22G, and
through the NASA Astrobiology Institute under Cooperative
Agreement No. CAN-02-OSS-02 issued through the Office of Space
Science.

\clearpage

\begin{deluxetable}{rrrrrr}
\tabletypesize{\footnotesize}
\tablecolumns{6} 
\tablewidth{0pc} 
\tablecaption{EGPs Used for Calculation of Mass Function} 
\tablehead{ 
\colhead{Planet Name} & \colhead{$e$}   & \colhead{$(M \sin i)/M_J$}    & \colhead{$a$ (AU)} & 
\colhead{Star Age (Gyr)}    & \colhead{Star Mass ($M_\odot$)} }
\startdata 
\colhead{}    &  \multicolumn{4}{c}{0.2 to $< 0.3$ $M_J$} \\ 
\cline{2-5} \\ 
HD 76700 b & 0.13 & 0.197 & 0.049 &  4.52 & 1.0 \\ 
HD 88133 b & 0.11 & 0.22 & 0.047 &  9.56 & 1.2 \\ 
HD 168746 b & 0.081 & 0.23 & 0.065 &  3.75 & 0.92 \\ 
HD 46375 b & 0.063 & 0.249 & 0.0398 &  4.96 & 1.0 \\ 
HD 109749 b & 0.01 & 0.28 & 0.0635 &  10.3 & 1.2 \\ 
\cline{2-5} \\ 
\colhead{}    &  \multicolumn{4}{c}{0.3 to $< 0.6$ $M_J$} \\ 
\cline{2-5} \\ 
HD 149026 b$^*$ & 0 &	0.36 & 0.042 & 5.8 & 1.3 \\
HD 63454 b & 0 & 0.385 & 0.0363 & 1.0 & 0.8 \\
HD 83443 b & 0.012 & 0.41 & 0.0406 & 2.94 & 0.79 \\
HD 75289 b & 0.034 & 0.42 & 0.046 & 4.96 & 1.05 \\
HD 212301 b & 0 & 0.45 & 0.0341 & 5.9 & 1.0 \\
51 Peg b & 0.013 & 0.472 & 0.0527 & 6.6 & 1.06 \\
HD 2638 b & 0 & 0.477 & 0.0436 & 3.0 & 0.93 \\
BD $-10$ 3166 b & 0.07 & 0.48 & 0.046 & 4.18 & 1.1 \\
HD 102195 b & 0.06 & 0.492 & 0.0491 & 2.4 & 0.93 \\
HD 187123 b & 0.023 & 0.528 & 0.0426 & 5.33 & 1.06 \\
OGLE-TR-111 b$^*$ & 0 & 0.53 & 0.047 & \nodata & \nodata \\
HAT--P-1 b$^*$& \nodata & 0.53 & 0.055 & 3.6 & 1.12 \\
OGLE-TR-10 b$^*$ & 0 & 0.54 & 0.04162 & \nodata & \nodata \\
\cline{2-5} \\ 
\colhead{}    &  \multicolumn{4}{c}{0.6 to $< 0.9$ $M_J$} \\ 
\cline{2-5} \\ 
$\upsilon$ And b & 0.012 & 0.69 & 0.059 & 2.41 & 1.3 \\
HD 209458 b$^*$ & 0 & 0.69 & 0.0474 & 4.72 & 1.01 \\
TrES-1 b$^*$ & 0 & 0.759 & 0.0394 & 2.41 & 0.87 \\
HD 330075 b & 0 & 0.76 & 0.043 & 6.21 & 0.95 \\
WASP-2 b$^*$ & \nodata & 0.88 & 0.0307 & \nodata & \nodata \\
WASP-1 b$^*$ & \nodata & 0.89 & 0.0382 & \nodata & \nodata \\
\cline{2-5} \\ 
\tablebreak
\colhead{}    &  \multicolumn{4}{c}{0.9 to $< 1.2$ $M_J$} \\ 
\cline{2-5} \\ 
XO-1 b$^*$ & \nodata & 0.9 & 0.0488 & \nodata & \nodata \\
HD 179949 b & 0.022 & 0.98 & 0.0443 & 2.05 & 1.24 \\
HD 188753A b & 0 & 1.14 & 0.0446 & \nodata & \nodata \\
HD 189733 b$^*$ & 0 & 1.15 & 0.0312 & 6.1 & 0.9 \\
OGLE-TR-113 b$^*$ & 0 & 1.19 & 0.0306 & \nodata & 1.35 \\
\cline{2-5} \\ 
\colhead{}    &  \multicolumn{4}{c}{1.2 to $< 1.5$ $M_J$} \\ 
\cline{2-5} \\ 
TrES-2 b$^*$ & 0 & 1.28 & 0.0367 & \nodata & 1.08 \\
OGLE-TR-56 b$^*$ & 0 & 1.29 & 0.0225 & 2 & 1.17 \\
OGLE-TR-132 b$^*$ & 0 & 1.32 & 0.0229 & $> 0.7$ & 0.78 \\
HD 149143 b & 0 & 1.33 & 0.0531 & 7.6 & 1.21 \\
HD 86081 b & 0.006 & 1.49 & 0.0346 & \nodata & 1.0 \\
\cline{2-5} \\ 
\colhead{}    &  \multicolumn{4}{c}{1.5 to $< 3$ $M_J$} \\ 
\cline{2-5} \\ 
HD 68988 b & 0.1249 & 1.86 & 0.0704 & 6.78 &1.2 \\
HD 73256 b & 0.029 & 1.87 & 0.0371 & 0.83 & 1.05 \\
HD 118203 b & 0.309 & 2.14 & 0.0703 & 4.6 & 1.23 \\
\cline{2-5} \\ 
\colhead{}    &  \multicolumn{4}{c}{3.0 to $< 4.5$ $M_J$} \\ 
\cline{2-5} \\ 
HIP 14810 b & 0.148 & 3.84 & 0.0692 & \nodata & 0.99 \\
$\tau$ Boo b & 0.023 &3.9 & 0.046 & 2.52 & 1.3 \\
\cline{1-6} \\
$^*$ transiting object &  &  &  &  & \\
\enddata
\end{deluxetable} 

\clearpage

\begin{figure}
\epsscale{.80}
\plotone{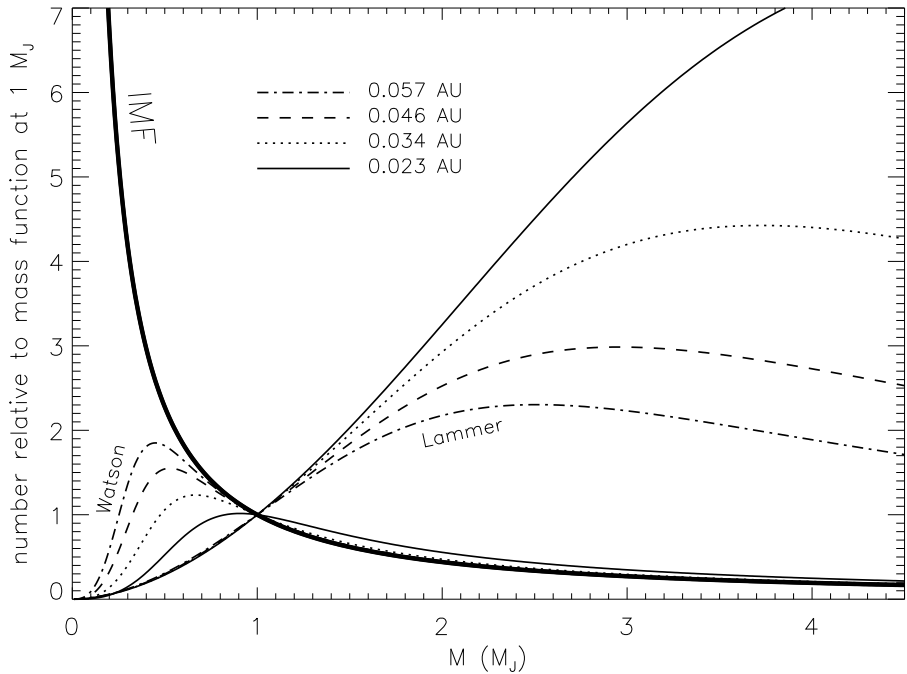}
\caption{Predicted mass functions $f(M,t)$ for EGPs suffering mass loss over 5 Gyr,
evaluated for four different orbital
radii, using the \citet{wat81} escape rate (left-hand curves with maxima at masses between
$0.4$ $M_J$ and $0.9$ $M_J$), and the two-orders-of-magnitude
larger \citet{lam03} escape rate (right-hand curves with maxima at
masses $\ge 2.3$ $M_J$).  The heavy curve (IMF) is the assumed mass distribution at the
start of mass loss.  All distributions are normalized to
unity at $M = 1 M_J$.
See text for a more detailed discussion of this plot.}
\end{figure}

\clearpage

\begin{figure}
\epsscale{.80}
\plotone{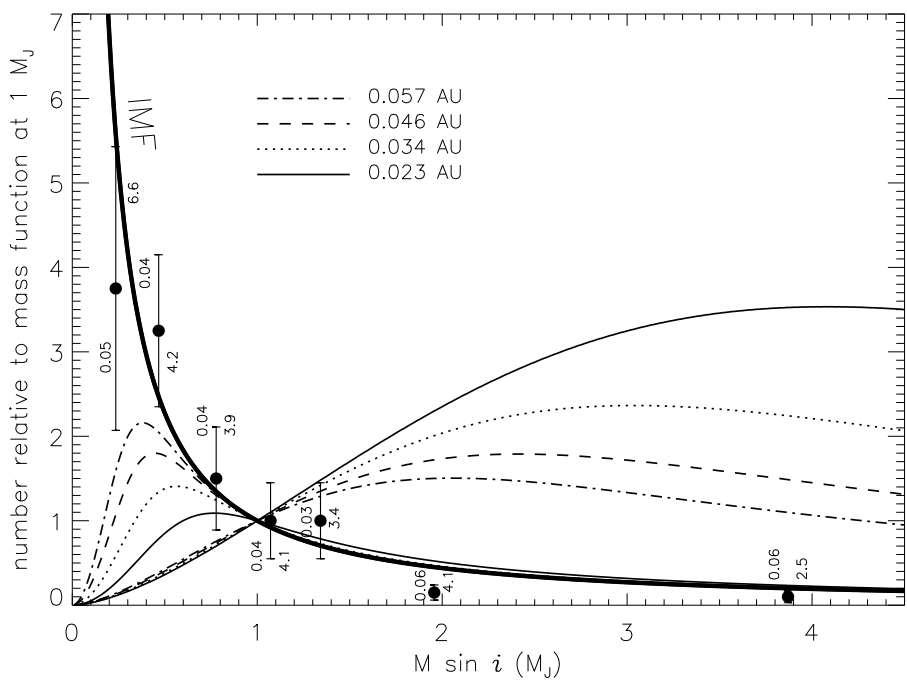}
\caption{Predicted mass functions $g(M \sin i, t)$
for EGPs suffering mass loss over 5 Gyr,
evaluated for the four different orbital
radii and the two mass-loss models, compared with
data for highly-irradiated EGPs. All distributions (theory and data) are normalized to
unity at $M \sin i = 1$ $M_J$.
See text for a more detailed discussion of this plot.}
\end{figure}

\end{document}